\documentclass[12pt]{article}
\catcode`\@=11

\@addtoreset{equation}{section}
\def\theequation{\arabic{equation}}
\def\theequation{\thesection\arabic{equation}}

\newcommand{\be}{\begin{equation}}
\newcommand{\ee}{\end{equation}}
\newcommand{\ba}{\begin{eqnarray}}
\newcommand{\ea}{\end{eqnarray}}

\def\NPB#1#2#3{{\it Nucl.~Phys.} {\bf{B#1}} (19#2) #3}
\def\PLB#1#2#3{{\it Phys.~Lett.} {\bf{B#1}} (19#2) #3}
\def\PRD#1#2#3{{\it Phys.~Rev.} {\bf{D#1}} (19#2) #3}

\def\JHEP#1#2#3{{\it J. High Energy Phys.} {\bf#1} (19#2) #3}

\def\part{\partial}

\def\simgt{\stackrel{>}{{}_\sim}}
\def\@normalsize{\@setsize\normalsize{15pt}\xiipt\@xiipt
\abovedisplayskip 14pt plus3pt minus3pt%
\belowdisplayskip \abovedisplayskip
\abovedisplayshortskip  \z@ plus3pt%
\belowdisplayshortskip  7pt plus3.5pt minus0pt}
\def\small{\@setsize\small{13.6pt}\xipt\@xipt
\abovedisplayskip 13pt plus3pt minus3pt%
\belowdisplayskip \abovedisplayskip
\abovedisplayshortskip  \z@ plus3pt%
\belowdisplayshortskip  7pt plus3.5pt minus0pt
\def\@listi{\parsep 4.5pt plus 2pt minus 1pt
            \itemsep \parsep
            \topsep 9pt plus 3pt minus 3pt}}

\def\underline#1{\relax\ifmmode\@@underline#1\else
        $\@@underline{\hbox{#1}}$\relax\fi}
\@twosidetrue
\relax

\catcode`@=12

\catcode`\@=11

\def\section{\@startsection{section}{1}{\z@}{3.5ex plus 1ex minus
   .2ex}{2.3ex plus .2ex}{\large\bf}}
\def\thesection{\arabic{section}.}
\def\thesubsection{\arabic{section}.\arabic{subsection}}

\def\ps@headings{\def\@oddfoot{}\def\@evenfoot{}
\def\@oddhead{\hbox{}\hfill
        \makebox[.5\textwidth]{\raggedright\ignorespaces --\thepage{}--
        \hfill }}
\def\@evenhead{\@oddhead}
\def\subsectionmark##1{\markboth{##1}{}} }
\renewcommand{\subsection}[1]{\addtocounter{subsection}{1}
\vspace{2.5mm}\par\noindent {\em \thesubsection . #1}\par
 \vspace{0.5mm} }
\ps@headings

\catcode`\@=12

\relax

\def\figcap{\section*{Figure Captions\markboth
        {FIGURECAPTIONS}{FIGURECAPTIONS}}\list
        {Fig. \arabic{enumi}:\hfill}{\settowidth\labelwidth{Fig. 999:}
        \leftmargin\labelwidth
        \advance\leftmargin\labelsep\usecounter{enumi}}}
 \relax
\def\tablecap{\section*{Table Captions\markboth
        {TABLECAPTIONS}{TABLECAPTIONS}}\list
        {Table \arabic{enumi}:\hfill}{\settowidth\labelwidth{Table 999:}
        \leftmargin\labelwidth
        \advance\leftmargin\labelsep\usecounter{enumi}}}
 \relax
\def\reflist{\section*{References\markboth
        {REFLIST}{REFLIST}}\list
        {[\arabic{enumi}]\hfill}{\settowidth\labelwidth{[999]}
        \leftmargin\labelwidth
        \advance\leftmargin\labelsep\usecounter{enumi}}}
 \relax

\catcode`\@=11

\def\marginnote#1{}
\newcount\hour
\newcount\minute
\newtoks\amorpm
\hour=\time\divide\hour by60
\minute=\time{\multiply\hour by60 \global\advance\minute by-
\hour}
\edef\standardtime{{\ifnum\hour<12 \global\amorpm={am}%
    \else\global\amorpm={pm}\advance\hour by-12 \fi
    \ifnum\hour=0 \hour=12 \fi
    \number\hour:\ifnum\minute<100\fi\number\minute\the\amorpm}}
\edef\militarytime{\number\hour:\ifnum\minute<100\fi\number\minute}
\def\draftlabel#1{{\@bsphack\if@filesw {\let\thepage\relax
  \xdef\@gtempa{\write\@auxout{\string
    \newlabel{#1}{{\@currentlabel}{\thepage}}}}}\@gtempa
    \if@nobreak \ifvmode\nobreak\fi\fi\fi\@esphack}
     \gdef\@eqnlabel{#1}}
\def\@eqnlabel{}
\def\@vacuum{}
\def\draftmarginnote#1{\marginpar{\raggedright\scriptsize\tt#1}}
\def\draft{\oddsidemargin -.5truein
        \def\@oddfoot{\sl preliminary draft \hfil
        \rm\thepage\hfil\sl\today\quad\militarytime}
        \let\@evenfoot\@oddfoot \overfullrule 3pt
        \let\label=\draftlabel
        \let\marginnote=\draftmarginnote
   
\def\@eqnnum{(\theequation)\rlap{\kern\marginparsep\tt\@eqnlabel}%
\global\let\@eqnlabel\@vacuum}  }
\def\preprint{\twocolumn\sloppy\flushbottom\parindent 1em
        \leftmargini 2em\leftmarginv .5em\leftmarginvi .5em
        \oddsidemargin -.5in    \evensidemargin -.5in
        \columnsep 15mm \footheight 0pt
        \textwidth 250mmin      \topmargin  -.4in
        \headheight 12pt \topskip .4in
        \textheight 175mm
        \footskip 0pt
        
\def\@oddhead{\thepage\hfil\addtocounter{page}{1}\thepage}
        \let\@evenhead\@oddhead \def\@oddfoot{} \def\@evenfoot{}  }
\def\titlepage{\@restonecolfalse\if@twocolumn\@restonecoltrue\onecolumn
     \else \newpage \fi \thispagestyle{empty}\c@page\z@
        \def\thefootnote{\fnsymbol{footnote}} }
\def\endtitlepage{\if@restonecol\twocolumn \else  \fi
        \def\thefootnote{\arabic{footnote}}
        \setcounter{footnote}{0}}  
\catcode`@=12
\relax

\def\ps@headings{\def\@oddfoot{}\def\@evenfoot{}
\def\@oddhead{\hbox{}\hfill
        \makebox[.5\textwidth]{\raggedright\ignorespaces --\thepage{}--
        \hfill }}
\def\@evenhead{\@oddhead}
\def\subsectionmark##1{\markboth{##1}{}} }

\ps@headings

\relax

\def\firstpage#1#2#3#4#5#6{
\begin{document}


\begin{titlepage}
\nopagebreak
\title{\begin{flushright}
        \vspace*{-1.8in}
        {\normalsize CPHT-PC748.1199}\\[-10mm]
       {\normalsize ROM2F-99/46}\\[-10mm]
        {\normalsize hep-th/9911205}\\[-4mm]
\end{flushright}
\vfill 
\vskip 4mm
{#3}}
\vskip -12mm
\author{ #4 \\[0.1cm] #5}
\maketitle
\vskip -9mm     
\nopagebreak 
\begin{abstract} {\noindent #6}
\end{abstract}
\vskip 36pt
\begin{center}
Based on talks presented by the authors at STRINGS '99, Potsdam,  
July 19-24  1999
\end{center}
\begin{flushleft}
\rule{16.1cm}{0.2mm}
\today
\end{flushleft}
\thispagestyle{empty}
\end{titlepage}}

\date{}
\firstpage{3118}{IC/95/34} {\large\bf Mass scales, supersymmetry  
breaking and open strings} 
{I. Antoniadis$^{\,a}$ \ and \ 
A. Sagnotti$^{\,b}$} 
{\small\sl
$^{a}$ Centre de Physique Th{\'e}orique (CNRS UMR 7644) \\[-5mm] 
\small\sl {\'E}cole Polytechnique \\[-5mm]
\small\sl {}F-91128 Palaiseau \ FRANCE \\
\small\sl $^{b}$ Dipartimento di Fisica \\[-5mm]
\small\sl Universit\`a di Roma ``Tor Vergata''\\[-5mm]
\small\sl INFN, Sezione di Roma ``Tor Vergata''\\[-5mm]
\small\sl 
Via della Ricerca Scientifica 1\\[-5mm]
\small\sl I-00133 Roma \ ITALY} 
{We review physical motivations and possible realizations 
of string vacua with large internal volume and/or low string scale and
discuss the issue of supersymmetry breaking. In particular, we describe the
key features of Scherk-Schwarz deformations in type I models and conclude
by reviewing the phenomenon of ``brane supersymmetry breaking'':  the tadpole
conditions of some type-I models require that supersymmetry be  {\it broken at
the string scale} on a collection of branes, while being  exact, to lowest
order, in the bulk and on other branes.}


\section{Introduction}

We have long been accustomed to accept that typical 
string effects, confined to very high scales, are beyond the reach of 
conceivable experiments. This state of affairs is directly implied by
the often implicit identification of the scale $M_s = (1/l_s)$ 
of string excitations with the Planck scale $M_p$, and 
has a rather stringent motivation in weakly-coupled heterotic
strings, that have been extensively studied during the last 
decade \cite{revhet}.
In this case, both gravitational and gauge effects originate from the
sphere topology, and the reduction on an internal space
of volume $V$ gives, in a self-explanatory notation, the four-dimensional 
effective action:
\be
S_H=\int d^4x{V\over\lambda_H^2}\, 
\left(\, l_H^{-8}{\cal R}\, + \, l_H^{-6}F^2 \, + \, \dots \, \right)\quad .
\label{SH}
\ee
This expression relates the four-dimensional Planck mass $M_p$ and the
four-dimensional gauge coupling $g$ at the string (unification) scale to the
heterotic string coupling $\lambda_H$ and to the heterotic string scale $M_H$,
according to
\be
M_H=g \, M_p \ , \qquad\qquad \lambda_H=g\, {\sqrt{V}\over l_H^3} \quad .
\label{het}
\ee
Thus, with the gauge coupling at unification determined by the
minimal supersymmetric Standard Model with the ``desert'' hypothesis, 
$g \sim 0.2$, one finds $M_H \sim 10^{18}$ GeV. Moreover, perturbative string
descriptions, with $\lambda_H < 1$, require internal 
volumes close to the string
size.\footnote{Up to T-dualities, we can always refer to compactification 
volumes larger than the string size.}

There are a few motivations, however, to consider string realizations
with small string scale and/or large extra dimensions \cite{a}. These were
already suggested by early studies of heterotic models 
although, as we have seen, they 
are outside the perturbative setting. Aside from the obvious 
phenomenological interest in meeting string effects at low energies, 
two main motivations have been associated over the years
with large
extra dimensions. The first was actually suggested by the estimated
value of the unification scale $M_{GUT}$, close enough to the string scale to
let one wonder whether the two should coincide. Taking into account
string threshold corrections \cite{kapl}, that grow linearly with internal
radii, one could try to ascribe the lack of coincidence 
to a geometric scale a couple or orders of magnitude below
$M_s$, at which new (Kaluza-Klein) physics would set in. 
This, however, would require generically that 
one of the internal radii be much larger than the others, with the
result of driving the heterotic description toward strong coupling,
as can be seen from eq. (\ref{het}).
Another motivation comes from supersymmetry breaking, that has long
been induced in heterotic models via a string extension of the 
Scherk-Schwarz mechanism \cite{ss,csss,ablt,a}. In Field Theory, the
Scherk-Schwarz  mechanism \cite{ss}
resorts to global symmetries to allow for more general mode expansions
in the internal space, and can be used, in principle, to induce 
supersymmetry breaking at arbitrary scales
if bosons and fermions are treated differently. In String Theory, however, 
the breaking scale
is necessarily $O(1/R)$, where $R$ is a typical geometric scale of the
internal space \cite{ablt}, and similar restrictions are met if the breaking
is induced by magnetic deformations \cite{costas}. Thus, taken at face value, 
these results imply that TeV-scale supersymmetry
breaking requires TeV-scale extra dimensions, with corresponding, nearly
accessible, Kaluza-Klein towers of excitations \cite{a,abq}.

With large extra dimensions, the effective field theory presents
inevitable subtleties. Most notably, at energies close to a geometric scale
$1/R$ related to a (large) extra dimension, the corresponding Kaluza-Klein 
tower starts contributing to renormalizations, and generically tends to drive
the system to strong coupling \cite{tv}, unless special conditions are met
\cite{a}. For the gauge couplings, this can be avoided if the
Kaluza-Klein towers fill $N=4$ multiplets, a condition nicely met in
interesting cases of string compactifications with no $N=2$ sectors
relative to the large compact coordinate \cite{a}.

With the advent of string dualities \cite{sd}, the heterotic string has 
lost its central role in the comparison with low-energy physics, while the 
problem of strong coupling can now be analyzed in a more
quantitative fashion, resorting to (dual) weakly-coupled descriptions 
provided by other
string models. Interestingly, these dual descriptions incorporate the
salient features, already envisaged from the perturbative heterotic corner, 
that are needed to grant a smooth behavior across the compactification scale. 
It is therefore 
instructive to explore the strong-coupling problem from the heterotic corner
with different numbers of large extra dimensions. In the next Section, we
thus present a brief review of the results of \cite{hetduals}, in particular
for what concerns type I models, and discuss
some issues raised by this analysis. For a more
detailed review, the interested reader may resort to \cite{iatrieste}. 
Our main motivation
here is to show how, in several of the resulting cases, the dual string models 
are characterized by low string scales, that at times can reach the TeV 
region, with additional interesting effects related to their 
towers of higher-spin excitations \cite{l}. 
In fact, as will become clear in the following, in any
string theory other than the heterotic, the simple relation (\ref{het}) that
fixes the string scale in terms of the Planck mass does not hold, and 
therefore the string tension becomes an arbitrary parameter
\cite{w}.\footnote{It was recently realized that the heterotic string scale
can also be lowered at weak coupling via small instantons \cite{bo}.} It can
be anywhere below the Planck scale, even 
at a few TeV. The main advantage
of having a string tension in the TeV region, 
aside from its obvious
experimental interest, is that it offers an automatic solution to the
gauge hierarchy problem, alternative to low-energy supersymmetry or technicolor
\cite{add}. Weakly coupled, low-scale strings
can be realized introducing either extra large {\it transverse} dimensions
felt only by gravitational interactions or an infinitesimal
string coupling. In the former case, the quantum gravity scale is also 
low, while gauge interactions are confined to lower-dimensional p-branes.
In the latter case, gravitational and string interactions remain suppressed by
the four-dimensional Planck mass. There is one exception to this general rule,
allowing for large longitudinal dimensions without a low string scale, when the
Standard Model is embedded in a six-dimensional fixed-point theory described
by a tensionless string \cite{hetduals}.

The very issue of supersymmetry breaking needs to be reconsidered in low-scale 
string models. With the string scale in the TeV region, the
original motivation for low-energy supersymmetry is apparently lost,
since in non-supersymmetric string vacua the bulk vacuum 
energy is typically determined by the string scale. While this
is a very favorable
state of affairs for the bulk spectrum, in type I models 
the  cosmological constant induced on our world-brane is actually
enhanced by the
volume of the transverse space. As a result, it is typically 
larger than the string scale, and tends to destabilize the hierarchy 
that one tries to enforce.

In Section 3 we discuss supersymmetry breaking 
by compactification in type I strings \cite{ads,kt,adds}. 
Referring to very simple models in nine dimensions, we show how the 
Scherk-Schwarz mechanism allows in this case two distinct possibilities,
according to whether the shifts are parallel or transverse with
respect to the resulting branes. In the latter case, they are
ineffective on their massless modes, that display an enhanced
supersymmetry. This is the phenomenon commonly denoted ``brane supersymmetry'',
first noted in \cite{ads} and developed further in \cite{kt,adds}. 
In the last Section, referring to a six-dimensional example, we 
review how the consistent definition of some type I models requires that,
to lowest order, supersymmetry be {\it broken at the string scale} on 
a collection of branes, while the bulk is unaffected \cite{bsb}. 
This phenomenon, we believe, is particularly 
intriguing. For the first time, supersymmetry breaking is not an option, 
but is {\it required} by the very consistency of a class of string models. 
The vacuum energy, restricted to the brane from which 
supersymmetry
breaking originates, is naturally protected against the 
destabilizing effects of gravitational radiative
corrections, since to lowest order the bulk is supersymmetric. 
Moreover, if our world is modeled 
resorting to this framework, the current 
experimental limits on short-distance gravitational effects \cite{subm} 
leave open the exciting
possibility of an (almost) exact supergravity a (sub)millimeter away
from it.

This is a joint version of the talks presented by the authors at STRINGS '99.
Transparencies and audio are available at \cite{strings99}.

\section{Large extra dimensions: the heterotic string and its duals}

\subsection{Type I strings and D-branes}

In ten dimensions, the strongly coupled $SO(32)$ heterotic string is 
dual to the type I string\footnote{T-dualities 
turn this model into the
type I$^\prime$ string, that in lower dimensions can also describe a class 
of M-theory compactifications.},
a theory where gravity
is described by unoriented closed
strings, while gauge interactions are described by unoriented open
strings whose ends are confined to D-branes. Therefore, in this
setting some of the six internal compact dimensions are 
longitudinal (parallel) and some are transverse to the D-branes. 
In particular, if the Standard Model were localized on a $p$-brane 
(with $p\ge 3$), there would be $p-3$ longitudinal and $9-p$ transverse 
compact dimensions. In contrast to the heterotic string, here 
gauge and gravitational interactions appear at
different orders of perturbation theory, and the corresponding effective
action reads
\be
S_{I}=\int d^{10}x \, \frac{1}{\lambda_I^2 l_I^8} \, {\cal R} + 
\int d^{p+1}x \, \frac{1}{\lambda_I l_I^{p-3}} \, F^2 \quad ,
\label{SI}
\ee
where the factor $1/\lambda_I$ in the gauge kinetic terms reflects their
origin from the disk diagram. 

Upon compactification to four dimensions, the Planck length and the gauge
couplings are given, to leading order, by
\begin{equation}
\frac{1}{l_P^2}=\frac{V_\parallel V_\perp}{\lambda_I^2 l_I^8}\quad ,\qquad
\frac{1}{g^2}=\frac{V_\parallel}{\lambda_I l_I^{p-3}}\quad ,
\label{I}
\end{equation}
where $V_\parallel$ ($V_\perp$) denotes the compactification volume 
longitudinal (transverse) to the $p$-brane. The second relation links
the weak coupling $\lambda_I<1$ to
sizes of the longitudinal space comparable to the
string length ($V_\parallel\sim l_I^{p-3}$), while the transverse volume
$V_\perp$ remains unrestricted. Combining eqs. (\ref{I}) gives
\begin{equation}
M_P^2=\frac{1}{g^4 v_\parallel}M_I^{2+n}R_\perp^n\ ,\qquad
\lambda_I =g^2 v_\parallel\ ,
\label{treei}
\end{equation}
to be compared with the heterotic relations (\ref{het}). Here 
$v_\parallel\simgt 1$ is the longitudinal volume in string units, 
and we are considering an isotropic transverse space with $n=9-p$ compact 
dimensions of radius $R_\perp$.

The relations (\ref{treei}) imply that the type I/I$^\prime$
string scale can be made hierarchically smaller than the Planck mass at
the expense of introducing extra large transverse dimensions that
interact only gravitationally \cite{add,st}. The weakness of
four-dimensional (4D) gravity
$M_I/M_P$ may then be attributed to the largeness of the transverse space
$R_\perp/l_I$. However, the (higher-dimensional) 
gravity becomes strong at the string scale, although the string coupling 
is weak, and indeed the
first of eq.(\ref{treei}) can be understood as a consequence of
the $(4+n)$-dimensional Gauss law for gravity, with
\be
G_N^{(4+n)}=g^4 l_I^{2+n}v_\parallel
\label{GN}
\ee
Newton's constant in $4+n$ dimensions.
Taking the type I string scale $M_I$ at 1 TeV,
one finds a size for the transverse dimensions $R_\perp$ varying from
$10^8$ km, to .1 mm (10$^{-3}$ eV), and down  to .1 fermi (10 MeV) for $n=1,2$,
or 6 large dimensions. Aside from the $n=1$
case, obviously excluded, all other cases are actually
consistent with observations, although barely for
$n=2$ \cite{add2}. In particular, sub-millimeter transverse directions are
compatible with  the present constraints from short-distance gravity
measurements, that have tested Newton's law only down to the cm \cite{subm}.

\subsection{Type IIA strings}

Upon compactification to 6 or fewer dimensions, the heterotic string
admits another dual description in terms of the 
type IIA string
compactified on a Calabi-Yau manifold. For simplicity, we
restrict ourselves to compactifications on $K3 \times T^2$, yielding $N=4$
supersymmetry, or more generally on Calabi-Yau manifolds that are $K3$
fibrations, yielding $N=2$ supersymmetry. In contrast to heterotic and type I
strings, non-abelian gauge symmetries in type IIA models arise
non-perturbatively (at arbitrarily weak coupling) in singular
compactifications, where the massless gauge bosons are provided by D2-branes
wrapped around (vanishing) non-trivial 2-cycles. The resulting gauge
interactions are localized on $K3$,
while matter multiplets arise from further singularities, and are 
completely localized in the 6D internal space.
As a result, the gauge kinetic terms are independent of the string coupling
$\lambda_{IIA}$, and the corresponding effective action is
\be
S_{IIA}=\int d^{10}x \frac{1}{\lambda_{IIA}^2 l_{IIA}^8} {\cal R} \, + \,
\int d^6 x {1\over l_{IIA}^2} F^2 \ ,
\label{SIIA}
\ee
to be compared with (\ref{SH}) and (\ref{SI}). 
Upon compactification to four dimensions,
for instance on a two-torus $T^2$, the gauge couplings are determined by
its size, $v_{T^2}$ in string units, while the Planck mass is controlled by
the 6D string coupling $\lambda_{6IIA}$:
\be
\frac{1}{g^2}=v_{T^2} \quad , \qquad
\frac{1}{l_P^2}=\frac{v_{T_2}}{\lambda_{6IIA}^2 l_{IIA}^2}
={1\over\lambda_{6IIA}^2}{1\over g^2l_{IIA}^2} \ .
\label{IIA}
\ee
The area of $T^2$ should therefore be of order $l_{IIA}^2$, while the
string scale is now related to the Planck mass according to
\be
M_{IIA}=g\lambda_{6IIA}M_P=
g\lambda_{IIA}M_P{l_{IIA}^2\over\sqrt{V_{K3}}}\, ,
\label{IIA2}
\ee
with $V_{K3}$ the volume of $K3$. Thus, in contrast to the type I relation
(\ref{treei}), only sensitive to the volume of the internal 
six-manifold, one now has the freedom to use both the string coupling and 
the $K3$ volume to
separate the Planck mass from a string scale at, say, 1 TeV. In particular, 
with a string-size internal manifold, an ultra-weak coupling
$\lambda_{IIA}=10^{-14}$ can account for the hierarchy between the electroweak
and Planck scales \cite{hetduals}. In this setting, 
despite the fact that the
string scale is so low, gravity remains weak up to the Planck scale,
while string
interactions are suppressed by the tiny string coupling, or equivalently
by the 4D Planck mass. Thus, no observable effects 
are left
for particle accelerators, aside from the production of KK excitations 
along the two TeV dimensions of $T^2$ felt by gauge interactions. 
Furthermore, the excitations
of the gauge multiplets have $N=4$ supersymmetry, even when $K3\times T^2$ is
replaced by a Calabi-Yau threefold that is a $K3$ fibration, while matter
multiplets are localized on the base (replacing the $T^2$) and have no KK
excitations, like the twisted states of heterotic orbifolds.

\subsection{Type IIB strings}

In type IIB constructions,
gauge symmetries still arise from vanishing
2-cycles of $K3$, but take the form of tensionless
strings in 6 dimensions, that originate from D3-branes wrapped on them. 
Only after a further $T^2$ reduction to four
dimensions does this theory reduce to an ordinary gauge theory, whose
coupling now involves the
shape (complex structure) $u_{T^2}$, rather than the volume $v_{T^2}$, 
of the torus. In this case one finds \cite{hetduals}
\be
\frac{1}{g^2}=u_{T^2}\quad ,\qquad
\frac{1}{l_P^2}=\frac{v_{T^2}}{\lambda_{6IIA}^2l_{IIB}^2}\quad ,
\label{IIB}
\ee
where,
for instance, for a rectangular torus the shape is the ratio of the
two radii,  $u_{T^2}=R_1/R_2$.
Comparing with eq. (\ref{IIA}), it is clear that the situation in type IIB is
the same as in type IIA, unless the size of $T^2$ is 
much larger than the string
length. Actually, since $T^2$ is felt by gauge interactions, its size 
cannot be larger than the TeV$^{-1}$, and thus the type IIB string scale 
should be much larger than the TeV. In particular, for a rectangular torus of 
radii $R$ and $g^2 R$
\be
M_{IIB}^2=g\lambda_{6IIB}{M_P\over R}\ ,
\label{IIB2}
\ee
so that the lowest value for the string scale, 
with a string coupling of order unity and $R\sim{\rm
TeV}^{-1}$, is $10^{11}$ GeV  \cite{hetduals}. This,
as we will see, is precisely
the case that describes the heterotic string with a single TeV dimension, and
is the only example of a weakly coupled theory with 
longitudinal dimensions larger than the string length. 
In the energy range between the KK scale $1/R$ and
the type IIB string scale, one has an effective 6D theory without gravity at a
non-trivial superconformal fixed point described by tensionless
strings, corresponding to D3-branes wrapped on the vanishing 2-cycles of a 
singular $K3$. Since the type IIB coupling is of order unity, gravity becomes
strong at the type IIB string scale, and the main 
experimental signatures at TeV
energies are in this case 
Kaluza-Klein excitations, as in type IIA models with tiny 
string coupling.

\subsection{Relation with the heterotic string}

As we mentioned previously, the type I/I$^\prime$ and type IIA/IIB theories
provide dual descriptions of the heterotic string at strong coupling.
Somewhat surprisingly, it turns out that all TeV scale
string models that we have discussed can be recovered as 
different strongly coupled decompactification limits, 
with only one large scale
in addition to the Planck-size heterotic tension.
More precisely, let us consider the heterotic string compactified on a
six-torus with $k$ large dimensions of radius $R\gg l_H$
and $6-k$ string-size dimensions. 
Applying the standard duality maps \cite{sd,hetduals}, it is simple
to show that the type I$^\prime$ theory with $n$ transverse dimensions 
provides
a weakly coupled dual description for the heterotic string only 
with $k=4,5,6$ large
dimensions, since otherwise the remaining T-dualities needed
to obtain volumes above the resulting 
string scale lead to strong coupling. 
$k=4$ is described by $n=2$, $k=6$ (for the $SO(32)$ gauge group) 
is described by $n=6$, while for $k=5$ one finds a type I$^\prime$ model with
five large and one extra-large transverse dimensions. 
The case $k=4$ is particularly interesting: the heterotic string with 4 large
dimensions at a TeV is described by a perturbative type I$^\prime$ theory
with the string scale at a TeV and gauge interactions confined to D7-branes
with two transverse dimensions of millimeter size,
T-dual to the two
string-size heterotic coordinates. On the other hand,
the type II theory provides a weakly
coupled description for $k=1,2,3,4$ and $k=6$ (for $E_8\times E_8$).
In particular, $k=1$ is described by type IIB with string tension at
intermediate energies, $k=2$ is
described by type IIA with tension and all compactification radii at
a TeV and an infinitesimal coupling $\lambda_{IIA}\sim l_H/R$, while for $k=3$
all four (transverse) $K3$ directions are extra large.

\section{Scherk-Schwarz deformations in type-I strings}

Scherk-Schwarz deformations can be introduced in type I strings 
following \cite{csss,ablt}, but present a few interesting
novelties, that may be conveniently exhibited referring to
a pair of 9D models \cite{ads}. To this end, we begin by
recalling that, for the type IIB string,
(the fermionic part of) the partition function can be written in the 
compact form
\be
{\cal T} = {|V_8 - S_8 |}^2  \label{ss1}
\ee
resorting to the level-one SO(8) characters
\ba
O_8 = { \vartheta_3^4 + \vartheta_4^4 \over 2 \eta^4} \quad , \quad
V_8 = { \vartheta_3^4 - \vartheta_4^4 \over 2 \eta^4} \quad , \nonumber \\
S_8 = { \vartheta_2^4 - \vartheta_1^4 \over 2 \eta^4} \quad , \quad
C_8 = { \vartheta_2^4 + \vartheta_1^4 \over 2 \eta^4} \quad , \label{ss2}
\ea
where the $\vartheta_i$ are Jacobi theta functions and $\eta$ is the
Dedekind function. In the usual toroidal reduction, where bosons and
fermions have the momentum modes
\be
p_L = \frac{m}{R} + \frac{n R}{\alpha^\prime} \ , \qquad
p_R = \frac{m}{R} - \frac{n R}{\alpha^\prime} \ , \label{ss3}
\ee
the 9D partition function is
\be
{\cal T} = {|V_8 - S_8 |}^2 \, Z_{mn} \  , \label{ss4}
\ee
where
\be
Z_{mn} \equiv \sum_{m,n} \frac{q^{{\alpha^\prime p_L^2 }/{4}} \
\bar{q}^{{\alpha^\prime p_R^2 }/{4}}}{ \eta \bar{\eta}} \ . \label{ss5}
\ee

A simple modification results in a Scherk-Schwarz breaking of 
space-time supersymmetry. There are actually two inequivalent choices, 
described by
\ba
{\cal T}_1 &=& Z_{m,2n} ( V_8 {\bar V}_8 + S_8 {\bar S}_8 ) +  
Z_{m,2n+1}( O_8
{\bar O}_8 + C_8 {\bar C}_8 )  \nonumber \\
&-& Z_{m+1/2,2n}( V_8 {\bar S}_8 + S_8 {\bar V}_8 ) -
Z_{m+1/2,2n+1}( O_8 {\bar C}_8 + C_8 {\bar O}_8 ) \label{ss6}
\ea
and
\ba
{\cal T}_2 &=& 
Z_{2m,n} ( V_8 {\bar V}_8 + S_8 {\bar S}_8 ) +  
Z_{2m+1,n}(O_8
{\bar O}_8 + C_8 {\bar C}_8 ) \nonumber \\
&-& Z_{2m,n+1/2}( V_8 {\bar S}_8 + S_8 {\bar V}_8 ) -
Z_{2m+1,n+1/2}( O_8 {\bar C}_8 + C_8 {\bar O}_8 ) \ , \label{ss7}
\ea
that may be associated to momentum or winding shifts of 
the usual fermionic modes ($V_8 \bar{S}_8$ and $S_8 \bar{V}_8$) 
relatively to the usual bosonic ones 
($V_8 \bar{V}_8$ and $S_8 \bar{S}_8$). 
The two choices are inequivalent, since T-duality 
along the circle interchanges type-IIB and 
type-IIA strings \cite{tduality}. Both deformed models have tachyon
instabilities at the scale of supersymmetry breaking for the low-lying
modes, $O(1/R)$ for the momentum deformation of eq. (\ref{ss5})
and $O(R/\alpha')$ for the winding deformation of eq. (\ref{ss6}).

The open descendants \cite{cargese} are essentially
determined by the choice of Klein-bottle projection ${\cal K}$ \cite{cc}, while
the other amplitudes ${\cal A}$ and ${\cal M}$ reflect the 
propagation of closed-string modes between boundaries and crosscaps.
In displaying the amplitudes of \cite{ads}, we implicitly confine our 
attention to internal radii such that (closed-string) tachyon instabilities 
are absent, and choose Chan-Paton assignments that remove them 
from the open sectors as well. We also impose some (inessential) 
NS-NS tadpoles, in order to bring the resulting expressions to their 
simplest forms.

Starting from the model of eq. (\ref{ss5}), corresponding to 
{\it momentum shifts}, the additional amplitudes are
\ba 
{\cal K}_1 &=& \frac{1}{2}  \ (V_8 - S_8) \ Z_m \ , \nonumber \\
{\cal A}_1 &=& \frac{n_1^2 + n_2^2}{2} ( V_8 Z_m - S_8 Z_{m + 1/2} )
+ n_1 n_2 ( V_8 Z_{m + 1/2} - S_8 Z_m ) \ ,  \nonumber \\ 
{\cal M}_1 &=& - \frac{ n_1 + n_2 }{2} ( {\hat
V}_8 Z_m - {\hat S}_8 Z_{m + 1/2} ) \ , \label{ss8}
\ea
while the tadpole conditions require that $n_1 + n_2 = 32$. Supersymmetry,
broken in the whole range $R > \sqrt{\alpha'}$,
is recovered asymptotically in the decompactification limit. This
vacuum, first described in \cite{bd}, is interesting in its own right,
since it describes the type I string at finite temperature (with Wilson
lines), but includes a 
rather conventional open spectrum, where bosonic and fermionic modes
have the usual $O(1/R)$ Scherk-Schwarz splittings of field-theory models. 

On the other hand, starting from the model of eq. (\ref{ss6}), 
corresponding to {\it winding shifts}, the additional amplitudes are \cite{ads}
\ba  
{\cal K}_2 &=& \frac{1}{2}  \ (V_8 - S_8) \ Z_{2m} + \frac{1}{2} \
(O_8 - C_8) \ Z_{2m+1} \ , \nonumber \\
{\cal A}_2 &=& \biggl( 
\frac{n_1^2 + n_2^2 + n_3^2 + n_4^2}{2} ( V_8 - S_8 )
+ ( n_1 n_3 + n_2 n_4 ) 
( O_8 - C_8 ) \biggr) Z_m \nonumber \\ &+& \biggl( (n_1 n_2 + 
n_3 n_4 ) (V_8  - S_8 )
+ (n_1 n_4 + n_2 n_3 ) (O_8 - C_8 )
\biggr) Z_{m+1/2} \ , \nonumber \\
{\cal M}_2 &=& - \frac{ n_1 + n_2 + n_3 + n_4 }{2} \, {\hat V}_8 \, Z_m +
\frac{ n_1 - n_2 - n_3 + n_4 }{2} \, {\hat S}_8 \, (-1)^m Z_m \ , 
\label{ss9}
\ea
while the tadpole conditions now require that $n_1 + n_2 = n_3 + n_4 = 16$.
Supersymmetry is recovered in the limit of vanishing 
$R$, where the whole tower of winding modes present in the 
vacuum-channel amplitudes collapses into additional 
tadpole conditions that eliminate $n_2$ and $n_3$. This is precisely 
the phenomenon of \cite{pw}, spelled out very clearly by these partition 
functions. The resulting open sector, described by
\ba
{\cal A}_2 &=& 
\frac{n_1^2 + n_4^2}{2} ( V_8 - S_8 ) Z_m + n_1 n_4 (O_8 - C_8 ) Z_{m+1/2} 
\ , \nonumber \\
{\cal M}_2 &=& - \frac{ n_1 + n_4 }{2} \, {\hat V}_8 \, Z_m +
\frac{ n_1 + n_4 }{2} \, {\hat S}_8 \, (-1)^m Z_m \ , 
\label{ss10}
\ea
has the suggestive gauge group $SO(16) \times SO(16)$, and
is rather peculiar. In the limit of small breaking $R$, aside from
the ultra-massive $(O,C)$ sector, it contains a conventional $(V,S)$
sector where supersymmetry, {\it exact} for the massless modes, is
broken {\it at the compactification scale} for the massive ones by the
unpairing of the corresponding Chan-Paton representations. This is
the phenomenon of ``brane supersymmetry'' that we alluded to in the
Introduction \cite{ads,kt,adds}, here present only for the massless modes.
However, as originally suggested in \cite{kt}, this setting can be
generalized to allow for entire open sectors with exact
supersymmetry, as in \cite{bg,adds2}. The arguments
of \cite{dg} can then connect, via a sequence of duality transformations, 
the $SO(16) \times SO(16)$ gauge group to the two Horava-Witten
walls \cite{hw} of M-theory, with the end result that this peculiar
breaking can be associated to an 11D Scherk-Schwarz deformation. We are
thus facing a simple perturbative description of a phenomenon whose
origin is non-perturbative on the heterotic side.
Several generalizations have been discussed, in six and four dimensions,
with partial or total breaking of supersymmetry \cite{ads,adds,
bg,adds2}.

After suitable T-dualities, these results can be put in 
a very suggestive form: while
the conventional Scherk-Schwarz breaking of ${\cal T}_1$ 
results from shifts {\it parallel} to a brane, the M-theory breaking 
of ${\cal T}_2$ results from shifts {\it orthogonal} to a 
brane, and is naturally 
ineffective on its massless modes.

\section{Brane supersymmetry breaking}
The last phenomenon that we would like to review in this talk, 
``brane supersymmetry breaking'' \cite{bsb}, provides an
answer to an old puzzle in the 
construction of open-string models where, in a number of interesting 
cases, the tadpole conditions have apparently no consistent solution 
\cite{mas}.
The simplest example is provided by the six-dimensional
$T^4/Z_2$ reduction where, as in \cite{cc}, the Klein-bottle projection
is reverted for all twisted states. In the 
resulting projected closed spectrum, described by
\ba
{\cal T} &=& \frac{1}{2} |Q_o + Q_v|^2 \Lambda + \frac{1}{2} |Q_o -
Q_v|^2 {\left|\frac{2 \eta}{\theta_2}\right|}^4 \nonumber \\
&+& \frac{1}{2} |Q_s +
Q_c|^2 {\left|\frac{2 \eta}{\theta_4}\right|}^4
+ \frac{1}{2} |Q_s +
Q_c|^2 {\left|\frac{2 \eta}{\theta_3}\right|}^4 \ , \nonumber \\
{\cal K} &=& \frac{1}{4} \left\{ ( Q_o + Q_v ) ( P + W ) -  2 \times
16 ( Q_s + Q_c ) \right\} \ , \label{bsb2}
\ea
the massless modes include
17 tensor multiplets and 4 hypermultiplets\footnote{A
quantized NS-NS $B_{ab}$ would lead to similar models with lower 
numbers tensor multiplets, that may be analyzed in a similar 
fashion \cite{carloB}.}. In writing eq. (\ref{bsb2}), where
$\Lambda$ is the whole Narain lattice sum while $P$ and $W$ 
are its restrictions to only momenta or windings, we have
resorted to the supersymmetric combinations of SO(4) characters 
\ba
Q_o &=& V_4 O_4 - C_4 C_4 \quad , \quad Q_v = O_4 V_4 - S_4 S_4 \ ,
\nonumber \\
Q_s &=& O_4 C_4 - S_4 O_4 \quad , \quad Q_c = V_4 S_4 - C_4 V_4  \ .
\label{bsb1}
\ea 
The reversal of the Klein-bottle projection for twisted states 
changes the relative sign of the crosscap contributions for N and D strings
or, equivalently, the relative charge of the O5 orientifold
planes relative to the O9 ones. This is clearly spelled out 
by the terms at the origin of the lattices,
\be
\tilde{\cal K}_0 = \frac{2^5}{4} \biggl\{ Q_o \biggl( \sqrt{v}  \pm
\frac{1}{\sqrt{v}}\biggr)^2 + Q_v \biggl( \sqrt{v}  \mp
\frac{1}{\sqrt{v}}\biggr)^2 \biggr\} \ , \label{bsb3}
\ee
where the upper signs refer to the standard choice, while the lower ones
refer to the reverted Klein bottle of eq. (\ref{bsb2}). In the
latter case one is forced to cancel a {\it negative} background O5 charge,
and this can be achieved introducing antibranes in the vacuum configuration.
The corresponding open sector \cite{bsb} 
results from a combination of D9 branes and D$\bar{5}$ 
antibranes, and involves the $N$ and $D$ charges 
and their orbifold breakings $R_N$ and $R_D$:
\ba
{\cal A} &=& \frac{1}{4} \biggl\{ (Q_o + Q_v) ( N^2 P  + D^2 W ) + 
2 N D (Q'_s + Q'_c) {\biggl(\frac{\eta}{\theta_4}\biggr)}^2  \label{a10}
\label{a13} \\
&+& (R_N^2 + R_D^2) (Q_o - Q_v) {\biggl(\frac{2
\eta}{\theta_2}\biggr)}^2 + 2 R_N R_D ( - O_4 S_4 - C_4 O_4 + V_4 C_4
+ S_4 V_4 ){\biggl(\frac{
\eta}{\theta_3}\biggr)}^2 \biggr\} \,  \nonumber \\
{\cal M} &=& - \frac{1}{4} \biggl\{ N P ( \hat{O}_4
\hat{V}_4  + \hat{V}_4 \hat{O}_4  - \hat{S}_4 \hat{S}_4
- \hat{C}_4 \hat{C}_4 ) -  D W ( \hat{O}_4
\hat{V}_4  + \hat{V}_4 \hat{O}_4  + \hat{S}_4 \hat{S}_4
+ \hat{C}_4 \hat{C}_4 ) \nonumber \\ &-&\!\!\!\!\! N( 
\hat{O}_4 \hat{V}_4 \!-\! \hat{V}_4 \hat{O}_4 \!-\! \hat{S}_4 \hat{S}_4
\!+\! \hat{C}_4 \hat{C}_4 )\left(
{2{\hat{\eta}}\over{\hat{\theta}}_2}\right)^2  \!\!+\! D( \hat{O}_4
\hat{V}_4 \!-\! \hat{V}_4 \hat{O}_4 \!+\! \hat{S}_4 \hat{S}_4
\!-\! \hat{C}_4 \hat{C}_4)\left(
{2{\hat{\eta}}\over{\hat{\theta}}_2}\right)^2  
\biggr\} \ . \nonumber
\ea
Supersymmetry is broken on the antibranes, and indeed the amplitudes 
involve the new 
characters $Q'_s$ and $Q'_c$, corresponding to a chirally flipped 
supercharge, that may
be obtained from eq. (\ref{bsb1}) upon the interchange of $S_4$ and $C_4$,
as well as other non-supersymmetric combinations.
The tadpole conditions determine the gauge group 
$[ SO(16) \times SO(16) ]_9 \times  
[ USp(16) \times USp(16) ]_{\bar{5}}$, and the
$99$ spectrum is supersymmetric, with (1,0) vector
multiplets for the $SO(16) \times SO(16)$ gauge group and a
hypermultiplet in the ${\bf\!
(16,16,1,1)}$.
On the other hand, the ${\bar 5} {\bar 5}$ spectrum is
non supersymmetric and,
aside from the $[ USp(16) \times USp(16) ]$ gauge vectors, 
contains quartets
of scalars in the ${\bf (1,1,16,16)}$, right-handed Weyl 
fermions in the
${\bf (1,1,120,1)}$ and ${\bf (1,1,1,120)}$ 
and left-handed Weyl fermions in 
the ${\bf (1,1,16,16)}$.
Finally, the ND sector, also non supersymmetric, comprises doublets of
scalars in the ${\bf (16,1,1,16)}$ and in the 
${\bf (1,16,16,1)}$, and additional
(symplectic) Majorana-Weyl fermions in the ${\bf
(16,1,16,1)}$ and ${\bf (1,16,1,16)}$. These fields are a peculiar feature
of six-dimensional space time, where the fundamental 
Weyl fermion, a pseudoreal spinor of $SU^*(4)$, 
can be subjected to a
Majorana condition if this is supplemented by the
conjugation in a pseudoreal representation.
All irreducible gauge and gravitational
anomalies cancel in this model, while the residual 
anomaly polynomial 
requires a generalized Green-Schwarz mechanism \cite{ggs} 
with couplings more general than those found in supersymmetric models.

It should be appreciated that the resulting
non-BPS configuration of branes and anti-branes has 
no tachyonic excitations, while the branes themselves experience no mutual
forces. Brane configurations of this type have received some attention lately
\cite{sen},
and form the basis of earlier constructions of non-supersymmetric type I
vacua \cite{nonsusy} and of their tachyon-free reductions \cite{carlo}.
As a result, the contributions to the vacuum energy, localized on
the antibranes, come solely from the M\"obius amplitude. 
The resulting
potential, determined by uncancelled D$\bar{5}$ NS-NS tadpole, is
\be
V_{\rm eff}=c{e^{-\phi_6}\over{\sqrt v}}=ce^{-\phi_{10}}
={c\over g_{\rm YM}^2}\ , \label{a18}
\ee
where $\phi_{10}$ is the 10D dilaton, that determines the Yang-Mills coupling
$g_{\rm YM}$ on the antibranes, and $c$ is a {\it positive} numerical 
constant. This potential (\ref{a18}) is clearly localized on the 
antibranes and positive, consistently with the
interpretation of this mechanism as global supersymmetry breaking.
One would also expect that, in the limit of vanishing 
D5 coupling, supersymmetry
be recovered, at least from the D9 viewpoint. While not true 
in six dimensions,
due to the peculiar chirality flip that we have described, the 
expectation is actually realized after compactification to four dimensions,
with suitable subgroups of the antibrane gauge group realized as internal
symmetries.

Several generalizations of this model have been discussed in \cite{bsb}.
Some include
tachyon-free combinations of branes and antibranes of the
same type, that extend
the construction of \cite{sug}. This more general setting has the
amusing feature of leading
to the effective stabilization of some geometric moduli, while some of
the resulting models, 
related to the $Z_3$ orientifold of \cite{abpss}, have interesting
three-family spectra of potential interest for phenomenology.

\vskip 24pt
\begin{flushleft}
{\large \bf Acknowledgments}
\end{flushleft}
The type I models discussed in this review were built with E. Dudas,
that we would like to thank for a very enjoyable collaboration. 
We are also grateful to C. Angelantonj and G. D'Appollonio, that have
contributed to several of these results.
This research was supported in part by the EEC under TMR contract 
ERBFMRX-CT96-0090.

\end{document}